\documentclass[aps,prd, 12pt, notitlepage,a4paper,floatfix,showpacs,nofootinbib,superscriptaddress,tikz]{revtex4-2}

\pdfoutput=1
\usepackage[usenames,dvipsnames]{color}  
\usepackage{graphicx} 

\usepackage[compat=1.1.0]{tikz-feynman}
\usepackage{comment}
\usepackage{xcolor}
\usepackage{bm}
\usepackage{bbold}
\usepackage{latexsym}
\usepackage{amsthm}
\usepackage{amsmath}
\usepackage{wrapfig}
\usepackage{amssymb}
\usepackage{cancel}
\usepackage{multirow}
\usepackage{pst-3d}                  
\usepackage{pst-grad}                
\usepackage{pst-node}                
\usepackage{pst-slpe}                
\usepackage{pst-plot}                %
\usepackage{pst-coil}                %
\usepackage{pst-math}                %
\usepackage{pstricks-add}            %
\usepackage{rotating}
\usepackage{geometry}
\usepackage{hhline}
\usepackage{amsmath}
\usepackage{float}
\usepackage{fancybox}
\usepackage[utf8]{inputenc}

\usepackage[colorlinks=true,citecolor=darkred,urlcolor=darkred, pdfborder={0 0 0}]{hyperref}
\usepackage[normalem]{ulem}

\usepackage{color,soul}
\setul{0.5ex}{0.3ex}
\setulcolor{blue}

\usepackage{braket}
\definecolor{linkcolor}{rgb}{0,0,0.5}
\allowdisplaybreaks
\allowdisplaybreaks[1]
\allowdisplaybreaks[2]

\definecolor{greenLinks}{rgb}{0, 0.6, 0}
\definecolor{blueLinks}{rgb}{0, 0, 0.6}
\definecolor{redLinks}{rgb}{0.6, 0, 0}
\definecolor{tempText}{rgb}{0.55, 0.10,0.67}
\definecolor{eprintLinks}{rgb}{0.4, 0.4, 0.4}
\definecolor{journalLinks}{rgb}{0.6, 0, 0}
\newcommand {\ignore}[1]{}

\usepackage{soul}
\definecolor{mightnightblue}{RGB}{25,25,112}

\definecolor{brown}{rgb}{0.59, 0.29, 0.0}

\definecolor{darkred}{rgb}{0.6,0,0}

\def\lsim{\mathrel{\rlap{\lower4pt\hbox{\hskip1pt$\sim$}}
    \raise1pt\hbox{$<$}}}
\def\gsim{\mathrel{\rlap{\lower4pt\hbox{\hskip1pt$\sim$}}
    \raise1pt\hbox{$>$}}}

 \newcommand{\AddrMPI}{Max-Planck-Institut f\"ur Kernphysik, Saupfercheckweg 1, 69117 Heidelberg, GERMANY}
 
  \newcommand{\AddrBhopal}{Department of Physics, Indian Institute of Science Education and Research - Bhopal, Bhopal Bypass Road, Bhauri, Bhopal 462066, INDIA}

\begin{document}

\title{CDF-II $W$ boson mass in the Dirac Scotogenic model}
\author{Salvador Centelles Chuli\'{a}}\email{chulia@mpi-hd.mpg.de}
\affiliation{\AddrMPI}
\author{Rahul Srivastava}\email{rahul@iiserb.ac.in}
\affiliation{\AddrBhopal}
\author{Sushant Yadav}\email{sushant20@iiserb.ac.in}
\affiliation{\AddrBhopal}

\begin{abstract}
\noindent

The Dirac scotogenic model provides an elegant mechanism which explains small Dirac neutrino masses and neutrino mixing, with a single symmetry simultaneously protecting the ``Diracness'' of the neutrinos and the stability of the dark matter candidate. Here we explore the phenomenological implications of the recent CDF-II measurement of the $W$ boson mass in the Dirac scotogenic framework. We show that, in the scenario where the dark matter is mainly a $SU(2)_L$ scalar doublet, it can satisfy all the theoretical and experimental constraints along with the CDF-II $W$ boson mass for the mass range of 58-86 GeV. However, unlike the Majorana scotogenic model, the Dirac version also has a ``dark sector'' $SU(2)_L$ singlet scalar. We show that if the singlet scalar is the lightest dark sector particle i.e. the dark matter then all neutrino physics and dark matter constraints along with the constraints from oblique  $S$, $T$ and $U$ parameters can be concurrently satisfied for $W$ boson mass in the CDF-II mass range, where the singlet dark matter mass is constrained up to around 500 GeV.
\end{abstract}

\maketitle

\section{Introduction}

The Standard Model (SM) of particle physics is the most precise theory in scientific history. Its multiple successes culminated with the discovery of a Higgs-like boson in 2012 \cite{CMS:2012qbp, ATLAS:2012yve}. However, we know that it must be an incomplete theory. The solid experimental/observational evidence for matter-antimatter asymmetry \cite{Planck:2018vyg}, dark matter (DM)\cite{Planck:2018vyg} and neutrino masses \cite{Kajita:2016cak, McDonald:2016ixn} cannot be accommodated in the SM framework. Moreover, other theoretical problems like the strong CP problem are present in the SM. If new scales higher than the electroweak (EW) scale are added to the theory in order to solve some or all of the previous issues then the hierarchy problem will arise.

Among the various shortcomings of SM, one of the most intriguing problems is the small but non-zero mass of neutrinos and the pattern of lepton mixing leading to neutrino oscillations. The unknown neutrino mass mechanism is tightly tied with the deeper question of neutrino nature. Neutrinos are singlets under the conserved symmetries of the SM, QCD and QED, and  therefore it is a natural expectation that the neutrino mass generation will break lepton number and thus neutrino will be of Majorana nature \cite{Minkowski:1977sc, Gell-Mann:1979vob, Schechter:1980gr, Schechter:1981cv}. However, new physics may very well introduce new symmetries that forbid these Majorana mass terms and neutrinos would instead be of Dirac type. This option is becoming popular in recent times, see \cite{CentellesChulia:2016rms,CentellesChulia:2016fxr,CentellesChulia:2017koy,Bonilla:2018ynb,CentellesChulia:2018gwr,CentellesChulia:2018bkz,CentellesChulia:2019xky,Ma:2014qra,Ma:2015raa,Ma:2015mjd,Peinado:2019mrn,Wang:2017mcy,Borah:2017leo,Jana:2019mgj,Jana:2019mez,Calle:2019mxn,Nanda:2019nqy,Gu:2019ird,Ma:2019byo,Ma:2019iwj,Correia:2019vbn,Saad:2019bqf,Ma:2019yfo,CentellesChulia:2020dfh,CentellesChulia:2020bnf,Guo:2020qin,delaVega:2020jcp,Borgohain:2020csn,Chulia:2021jgv,Bernal:2021ppq,Biswas:2021kio,Mahanta:2021plx,Hazarika:2022tlc, Mishra:2021ilq,Chowdhury:2022jde,Biswas:2022vkq,Li:2022chc,Leite:2020wjl} for a few selected examples. Detecting neutrinoless double beta decay would settle this issue \cite{Schechter:1981bd}, but so far the experimental searches remain inconclusive in this regard \cite{KamLAND-Zen:2016pfg}.


Any mass mechanism for neutrinos, whether Dirac or Majorana, will require the addition of new particles beyond SM. In general, these new particles can have an impact on the EW precision observables. It is convenient to parameterize these new effects, in particular on the $W$ and $Z$ gauge boson masses, in a model-independent way through the oblique  $S$, $T$ and $U$ parameters \cite{Peskin:1991sw}. Until very recently, the philosophy for model-building was that the new physics contributions should be small enough so that they do not lead to unacceptably large corrections in the EW sector. However, the theoretical expectation should be to eventually observe deviations from the SM predictions, since we know that the SM has to be extended in some way. Indeed, in 2022 the CDF-II collaboration reported a $7\sigma$ excess on the mass of the $W$ boson~\cite{CDF:2022hxs} with respect to the SM prediction~\cite{ParticleDataGroup:2020ssz}. This experimental result is in direct clash with ATLAS measurements of the $W$ boson mass \cite{DiMarco:2019ppm}. This issue can be hopefully clarified as the new run of LHC should bring new data very soon.  In any case, if CDF-II measurements are correct then new physics beyond Standard Model (BSM) is definitely needed. Keeping the CDF-II results in mind, the  new fits to the EW precision oblique parameters $S$, $T$ and $U$ including the CDF-II $W$ boson mass measurement have been performed \cite{Flacher:2008zq, Asadi:2022xiy,deBlas:2022hdk, Lu:2022bgw, Balkin:2022glu, Paul:2022dds, Gu:2022htv, Kawamura:2022uft, Strumia:2022qkt}, again indicating the need for BSM physics corrections to the $W$ boson mass.
As emphasized before, these sorts of deviations from the SM predictions are actually expected if the SM is extended in order to explain some of its shortcomings. The class of models that can explain the $m_W$ anomaly is, therefore, very wide. See 
\cite{Dcruz:2022dao,Thomas:2022gib,Yang:2022gvz,Yuan:2022cpw,Chowdhury:2022dps,Barman:2022qix,Wang:2022dte,Cai:2022cti,Cheng:2022aau,Athron:2022qpo,Liu:2022vgo,Ma:2022emu, Gao:2022wxk,Gupta:2022lrt,Tan:2022bip,Addazi:2022fbj,Li:2022gwc,Babu:2022pdn,Kim:2022xuo,Kim:2022hvh,Zhou:2022cql,Babu:2022pdn,Abouabid:2022lpg,Fan:2022dck,Zhu:2022tpr,Kawamura:2022fhm, Fonseca:2022wtz,Biekotter:2022abc,Du:2022brr, Endo:2022kiw, Heo:2022dey, Han:2022juu, Ahn:2022xeq, FileviezPerez:2022lxp, Ghoshal:2022vzo, Kawamura:2022fhm, Kawamura:2022uft, Kanemura:2022ahw, Nagao:2022oin, Zhang:2022nnh, Popov:2022ldh, Arcadi:2022dmt, Chowdhury:2022moc, Borah:2022obi, Cirigliano:2022qdm, Zeng:2022lkk, Du:2022fqv, Ghorbani:2022vtv, Bhaskar:2022vgk, Baek:2022agi, Cao:2022mif, Borah:2022zim, Batra:2022org, Lee:2022gyf,Bahl:2022xzi, Batra:2022pej, Chakrabarty:2022voz} for a few examples.

Among these, simple models which can simultaneously explain the existence of DM and small neutrino masses are of particular interest. A simple but well-motivated extension of the SM in which neutrinos obtain mass via a one-loop diagram is the so-called scotogenic model \cite{Ma:2006km}. In its canonical version a new 'ad-hoc' $Z_2$  symmetry is added to the model in order to stabilize the lightest particle in the loop, which then becomes a viable DM candidate. 
If neutrinos are Dirac, however, it is possible to obtain the same stability using just a chiral $U(1)_{B-L}$ symmetry. This $U(1)_{B-L}$ is multipurpose: it stabilizes the DM, protects the Diracness of neutrinos and forbids the tree-level neutrino mass term, thus naturally explaining the smallness of neutrino mass \cite{Bonilla:2018ynb}. An additional consequence of the chiral $U(1)_{B-L}$ is that the lightest neutrino will be massless. See \cite{Guo:2020qin, Hazarika:2022tlc} for other phenomenological analyses of the Dirac scotogenic variants and \cite{Li:2022chc} for a study on its relation with B anomalies.
Here, we will focus on the connection between DM and neutrino masses in the particular framework of the Dirac scotogenic model \cite{Bonilla:2018ynb}, the implications of the new CDF-II result and the difference with respect to the Majorana scotogenic model. 

The work is structured as follows. In Sec.~\ref{sec:model} we present the basic details of the model. In Sec.~\ref{sec:numass} we explore the neutrino mass and mixing of the model, with particular focus on the parameter $\kappa$, the dimensionful analogue of the $\lambda_5$ parameter in the original scotogenic model. In Sec.~\ref{sec:Wmass} we compute the model corrections to the $W$ boson mass and show that the pure $SU(2)_L$ doublet scalar DM case is severely constrained from the need to satisfy CDF-II $W$ boson mass measurement while at the same time keeping the oblique $S$, $T$ and $U$ parameters within their global fit range. Finally, in Sec.~\ref{sec:DM} we analyze the DM sector in the two extreme cases, the doublet scalar DM and the singlet scalar DM. We find that the doublet DM case is compatible with the relic abundance and direct detection constraints only if its mass is in the range of 58-86 GeV. The singlet scalar DM can also satisfy all the constraints simultaneously. We conclude with final remarks in Sec.~\ref{sec:conclusions}

\section{The model setup}
\label{sec:model}

We will now flesh out the most important characteristics of the model presented in \cite{Bonilla:2018ynb}. The field and symmetry inventory is given in Tab.~\ref{tab:model}. The key ingredient of the model is the chiral anomaly free $B-L$ symmetry\footnote{Being anomaly free, the $U(1)_{B-L}$ symmetry can be gauged. However, here for sake of simplicity, we will take it to be a global symmetry.}\cite{Ma:2014qra, Ma:2015raa, Ma:2015mjd} with a triple role: protecting the stability of DM, protecting the  Dirac nature of the neutrinos and the smallness of neutrino masses by forbidding the tree-level coupling with the Higgs field. It also predicts that the lightest neutrino is massless. The $B-L$ symmetry is then broken to a residual $Z_n$, $n \in$ positive integer; residual subgroup which remains unbroken. Depending on the $Z_n$ subgroup and charges of the lepton doublets under it, the neutrinos can be Dirac or Majorana in nature \cite{Hirsch:2017col, Bonilla:2018ynb}. Here, we will focus on the $U(1)_{B-L} \to Z_6$ residual subgroup which arises naturally if $B-L$ symmetry is broken in units of three. The residual $Z_6$ then ensures the Dirac nature of neutrinos by forbidding all possible Majorana terms to all loop orders of perturbation. It simultaneously also protects the stability of the lightest ``dark sector'' particle which can be a viable DM candidate \cite{Bonilla:2018ynb}.

\begin{table}
\begin{center}
\begin{tabular}{| c || c | c  || c | c |}
  \hline 
& \hspace{0.1cm}  Fields  \hspace{0.1cm}          &  \hspace{0.1cm}  $SU(2)_L \otimes U(1)_Y$  \hspace{0.1cm} & \hspace{0.1cm} $U(1)_{B-L}$  \hspace{0.1cm}  & \hspace{0.15cm} {\color{blue}$\mathcal{Z}_{6}$}  \hspace{0.15cm}              \\
\hline \hline
\multirow{4}{*}{ \begin{turn}{90} \hspace{-0.15cm} Fermions \end{turn} } &
 $L_i$        	  &    ($\mathbf{2}, {-1/2}$)    & $-1$ 	  &	 {\color{blue}$\omega^4$}                \\	
&   $\nu_{R_i}$       &   ($\mathbf{1}, {0}$)    & $(-4, -4, 5)$  &  	 {\color{blue}$\omega^4$} \\
&   $N_{L_l}$    	  &   ($\mathbf{1}, {0}$)    & $-1/2$   &    {\color{blue} $\omega^5$}     \\
&  $N_{R_l}$     	  &  ($\mathbf{1}, {0}$) 	 & $-1/2$    &  {\color{blue}$\omega^5$}    \\
\hline \hline
\multirow{4}{*}{ \begin{turn}{90} \hspace{0.2cm} Scalars \end{turn} } &
 $H$  		 &  ($\mathbf{2}, {1/2}$)            & $0$    & {\color{blue} $1$}   \\
& $\eta$          	 &  ($\mathbf{2}, {1/2}$)    & $1/2$      &  {\color{blue}$\omega$}     \\
& $\xi$             &  ($\mathbf{1}, {0}$)       & $7/2$     &	{\color{blue}$\omega$} \\	
    \hline
  \end{tabular}
\end{center}
\caption{Charge assignment for all the fields. $\mathcal{Z}_6$ is a residual symmetry coming from $U(1)_{B-L}$ breaking, see \cite{Bonilla:2018ynb} for a detailed discussion. $\omega$ is the sixth root of unity, i.e. $\omega^6=1$.}
 \label{tab:model} 
\end{table}

In Tab. \ref{tab:model} apart from SM particles, we have added three right-handed neutrinos $\nu_{R_i}$; $i = 1,2,3$ with $B-L$ charges of $(-4,-4,5)$ and dark sector particles: the singlet fermions $N_{L_i}, N_{R_i}$; $i = 1,2,3$, the $SU(2)_L$ doublet scalar $\eta$ and the $SU(2)_L$ singlet scalar $\xi$, with $B-L$ charges as shown in Tab. \ref{tab:model}. As mentioned before, the lightest dark sector particle will be stable and will be the DM candidate in our model.

\begin{figure}[th]
\includegraphics[width=10cm]{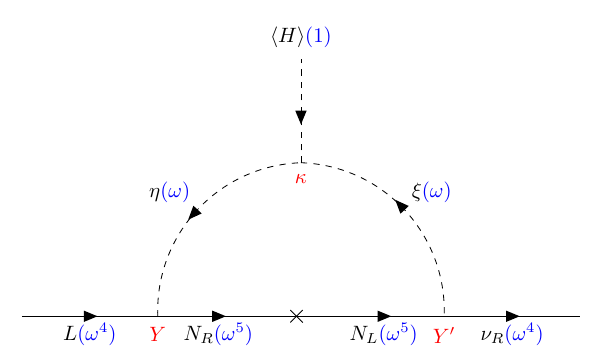}
        \caption{\begin{footnotesize}Leading neutrino mass generation diagram. The tree-level mass term between left and right-handed neutrinos is forbidden by the chiral $U(1)_{B-L}$ symmetry, only appearing at the one-loop level after the $U(1)_{B-L} \to Z_6$ symmetry breaking. The unbroken residual $Z_6$ symmetry simultaneously ensures that neutrinos remain Dirac in nature and the lightest dark sector particle is completely stable. \end{footnotesize}}
        \label{fig:numass}
\end{figure}

The relevant Yukawa Lagrangian for neutrino masses is given by

\begin{equation}
    -\mathcal{L}_{Y}\supset Y_{i j} \Bar{L}_{i} \tilde{\eta} N_{R_{j}}+Y_{i k}^{\prime} \bar{N}_{L_{i}} \nu_{R_{k}} \xi+M_{i j} \bar{N}_{R_{i}} N_{L_{j}}+h . c.
    \label{eq:yukawas}
\end{equation}

where $\tilde{\eta} = \iota \tau_{2} \eta^{*}$, $\tau_{2}$ is the $2^{nd}$ Pauli matrix and the indices $i,j \in \{1,2,3\}$, $k \in \{1, 2\}$. \\
$Y$ and $Y^{\prime}$ are dimensionless Yukawa couplings, while $M$ is a Dirac mass term for the fermions $N_L$ and $N_R$. While $Y$ and $M$ are $3\times 3$ matrices, $Y^{\prime}$ is instead $3 \times 2$ owing to the $B-L$ charges of the right handed neutrino $\nu_R$. The total neutrino mass matrix is, therefore, rank $2$ and one neutrino is massless. \\
Eq.~\ref{eq:yukawas} generates a one-loop neutrino mass in the Dirac version of the scotogenic model. See Fig.~\ref{fig:numass}. Given the different $B-L$ charges for $L_i$ and $\nu_{R_i}$, the tree-level direct coupling for neutrinos is forbidden and the neutrino masses can only be generated at one-loop level after $U(1)_{B-L} \to Z_6$ breaking. 

In the scalar sector, the Higgs field $H$ will be SM-like. Owing to their $Z_6$ symmetry charges, the scalar fields $\eta$ and $\xi$ will be complex fields, thus the real and imaginary parts will be degenerate in mass. However, the neutral component of $\eta$ can mix with the singlet $\xi$. The mixing between the components of $H$ and $\eta, \xi$ is forbidden by the $Z_6$ symmetry. The general form of the scalar potential is given by

\begin{eqnarray}
V & = & -\mu_{H}^{2} H^{\dagger} H +\mu_{\eta}^{2} \eta^{\dagger} \eta + \mu_{\xi}^{2} \xi^{*} \xi
+\frac{1}{2} \lambda_{1}(H^{\dagger} H)^{2}+\frac{1}{2} \lambda_{2}(\eta^{\dagger} \eta)^{2}+\frac{1}{2} \lambda_{3}(\xi^{*} \xi)^{2} + \lambda_{4}(H^{\dagger} H)(\eta^{\dagger} \eta) \nonumber \\
& + &  \lambda_{6}(\eta^{\dagger} \eta)(\xi^{*} \xi)+\lambda_{7}(H^{\dagger} \eta)(\eta^{\dagger} H)+ \lambda_8(H^{\dagger} H)(\xi^{*} \xi) + (\kappa \, \eta^{\dagger} H \xi + h.c.)
\label{eq:pot}
\end{eqnarray}

Here we have avoided the use of $\lambda_5$ in order to prevent confusion with the $\lambda_5$ parameter of the Majorana scotogenic case. The $U(1)_{B-L} \to Z_6$ breaking happens because of the presence of the soft term $\kappa \, \eta^{\dagger} H \xi + h.c$. Thus, in the Dirac scotogenic model, $\kappa$ plays the role that $\lambda_5$ plays in the Majorana version. 
%
The requirement to have a stable minimum for the potential implies the following conditions\cite{Kannike:2016fmd}

\begin{eqnarray}
& & \lambda_{1}, \lambda_{2}, \lambda_{3}  >  0; \hspace{0.5cm} \lambda_{4} > -\sqrt{\lambda_{1}\lambda_{2}}, \hspace{0.5cm} \lambda_{6} > -\sqrt{\lambda_{2}\lambda_{3}}, \hspace{0.5cm} \lambda_8 > -\sqrt{\lambda_{1}\lambda_{3}}, \nonumber\\
& & \sqrt{\frac{\lambda_{3}}{2}}\lambda_{4}  +  \sqrt{\frac{\lambda_{1}}{2}}\lambda_{6} + \sqrt{\frac{\lambda_{2}}{2}}\lambda_8 + \sqrt{\frac{\lambda_{1}\lambda_{2}\lambda_{3}}{8}} > -\sqrt{(\lambda_{4} +\sqrt{\lambda_{1}\lambda_{2}})(\lambda_8+\sqrt{\lambda_{1}\lambda_{3}})(\lambda_{6}+\sqrt{\lambda_{2}\lambda_{3}})} \hspace{0.3cm} .
\label{eq:vacuumstability}
  \end{eqnarray}
  
Fleshing out the $SU(2)_L$ components of the scalars, we can write the doublets as
\begin{eqnarray}
& & \begin{aligned}
  H = \begin{pmatrix}
H^+\\
H^0
\end{pmatrix}, & \hspace{1cm}& \eta = \begin{pmatrix}
\eta^+\\
\eta^0
\end{pmatrix}
\end{aligned}\\
   & & H^0=\frac{1}{\sqrt{2}}(v+h+iA), \hspace{0.5cm} \eta^0=\frac{1}{\sqrt{2}}(\eta_{R}+i\eta_{I}), \hspace{0.5cm} \xi=\frac{1}{\sqrt{2}}(\xi_{R}+i\xi_{I}) \hspace{0.2cm}.
\end{eqnarray}

However, as emphasized before, the residual symmetry $Z_6$ will force the neutral imaginary and real components of each dark sector scalar to be degenerate. We can now compute the masses of the physical scalar states after symmetry breaking 
\begin{equation}
    m_{h}^{2} = \lambda_{1}v^2,
\end{equation}
\begin{equation}
    m_{\eta^{\pm}}^{2} = \mu_{\eta}^{2} + \frac{\lambda_{4}}{2}v^2,
    \label{eq:metaplus}
\end{equation}

The real part of $\xi$ will mix with the real part of $\eta^0$ and similarly, the imaginary part of $\xi$ will mix with the imaginary part of $\eta^0$ with the same mixing matrix.

\begin{equation}
    m_{(\xi_{R},\eta_{R})}^{2}=m_{(\xi_{I},\eta_{I})}^{2}=\begin{pmatrix}
    \mu_{\xi}^{2} + \lambda_8 \frac{v^2}{2} & \kappa \frac{v}{\sqrt{2}}\\
    \kappa \frac{v}{\sqrt{2}} & \mu_{\eta}^{2} + (\lambda_{4}+\lambda_{7}) \frac{v^2}{2}
    \end{pmatrix}
    \label{eq:metamix}
\end{equation}
We can compute the mixing angle
\begin{equation}
    \tan2\theta=\frac{\sqrt{2}\kappa \, v}{ (\mu_{\xi}^{2}-\mu_{\eta}^{2})+(\lambda_8-\lambda_{4}-\lambda_{7})\frac{v^2}{2}}
    \label{eq:mix}
\end{equation}

and the mass eigenstates for the real/imaginary part of neutral scalars $\eta^0$ and $\xi$ are given by
\begin{eqnarray}
    m^2_{1R}=m^2_{1I}= \left(\mu_{\xi}^{2} + \lambda_8 \frac{v^2}{2}\right) \cos^2{\theta} + \left(\mu_{\eta}^{2} + (\lambda_{4}+\lambda_{7}) \frac{v^2}{2}\right)\sin^2{\theta} - 2\kappa v \sin{\theta} \cos{\theta}=m^2_{\xi}\\
m^2_{2R}=m^2_{2I}= \left(\mu_{\xi}^{2} + \lambda_8 \frac{v^2}{2}\right) \sin^2{\theta} + \left(\mu_{\eta}^{2} + (\lambda_{4}+\lambda_{7}) \frac{v^2}{2}\right)\cos^2{\theta} + 2\kappa v \sin{\theta} \cos{\theta}=m^2_{\eta^0}
\end{eqnarray}

Finally, the tree level perturbativity\footnote{See \cite{Krauss:2017xpj} for a more detailed analysis on loop-corrected perturbativity conditions.} of the dimensionless couplings additionally implies

\begin{equation}
    Tr(Y^{\dagger} Y ) < 4\pi, \hspace{0.2cm} Tr(Y^{\prime \dagger} Y^{\prime} ) < 4\pi, \hspace{0.2cm}  |\lambda_j| \leq \sqrt 4 \pi
    \label{eq:perturbativity}
\end{equation}

where $Y$ and $Y^{\prime}$ are the Yukawa couplings of Eq.~\ref{eq:yukawas} while $\lambda_j$ are the scalar potential quartic couplings of Eq.~\ref{eq:pot}, where $j\in \{2,3,4,6,7,8\}$.

In summary, apart from a vev carrying $SU(2)_L$ doublet scalar $H$, we have two additional scalars $\eta$ and $\xi$ both of which belong to the dark sector. The lighter neutral mass eigenstate can be a good candidate for DM if it is the lightest dark sector particle with its stability ensured by the unbroken residual $Z_6$ symmetry.


\section{Neutrino masses}
\label{sec:numass}

The neutrino mass generation mechanism in this model is independent of the modifications to the electroweak precision parameters. Therefore, the results of previous works apply here too. However, we will flesh out the neutrino properties in the model because our setup is more restricted since the lightest neutrino is massless.
The reason for its masslessness comes from the chiral $B-L$ structure shown in \cite{Bonilla:2018ynb}. We can calculate neutrino masses from the diagram Fig.~\ref{fig:numass} as

 \begin{equation}
    (M_\nu)_{ij} = \frac{1}{16\pi^2} \sum\limits_{k=1}^3 Y_{ik} Y'_{kj} \frac{\kappa v}{m^2_\xi-m^2_\eta} M_k \sum\limits_{l=1}^2 (-1)^l B_0(0,m_l^2,M_k^2).
\end{equation}

Where $m_l, M_k$ are the masses of the light and heavy neutrinos, respectively, $Y$, $Y'$ and $\kappa$ are the couplings described in Eq.~\ref{eq:yukawas} and Fig.~\ref{fig:numass}, $m_\xi$ and $m_{\eta^0}$ are the neutral scalar mass eigenvalues, $v$ is the SM vev and $B_0$ is the Passarino-Veltman function \cite{Passarino:1978jh}. It is important to mention that this setup leads to only two massive neutrinos since $Y'$ is a rank-2 matrix.

Since the Yukawa matrices are free parameters it is clear that it is possible to fit the mixing parameters of neutrino oscillations given by the global fit \cite{deSalas:2020pgw}. In the analysis that follows we will impose neutrino masses and mixing inside the $3\sigma$ range of the global fit. As a benchmark scenario, we have assumed the inverted ordering of the light neutrino masses. This is a natural choice since the lightest neutrino mass is zero. However, assuming normal ordering will not change the conclusions.

In the scalar DM case, we can take the fermionic mass $M_k \gg m_\xi, m_{\eta^0}$. In this case, the neutrino mass scale is approximately given by

\begin{equation}
   m_\nu \sim \frac{1}{16\pi^2} \frac{\kappa v}{M} Y Y'
\end{equation}

For example, for Yukawas of order $1$, a fermionic mass of $500$ GeV and $m_\nu < 0.1$ eV leads to a value of $\kappa < 30$ GeV. It is important to mention that $\kappa$ is the dimensionful equivalent of $\lambda_5$ in the traditional Majorana scotogenic model. In the limit, $\kappa \to 0$, the symmetry of the Lagrangian gets enhanced from a $Z_6$ to a $U(1)_{B-L}$ symmetry \cite{Bonilla:2018ynb} and its smallness can be associated with an inverse seesaw mechanism \cite{CentellesChulia:2020dfh}. Thus, in the same spirit as in \cite{Ma:2006km}, $\kappa$ is a symmetry-protected parameter which can therefore be small. For these reasons, we will consider the limit $\kappa \ll \Lambda_{EW}$ throughout this work. One could deviate from this limit by taking heavier neutral fermions or smaller Yukawas.
Working in the small $\kappa$ regime has an additional consequence in the scalar sector. As can be seen from Eq.~\ref{eq:mix}, in this limit there won't be a sizeable mixing between the scalars, with very important consequences in the DM sector and mass of $W$ boson $m_W$ as we will now discuss.

\section{$W$ mass and the $S$, $T$, $U$ parameters}
 \label{sec:Wmass}

\begin{figure}[th]
    \centering
    \includegraphics[width=10cm]{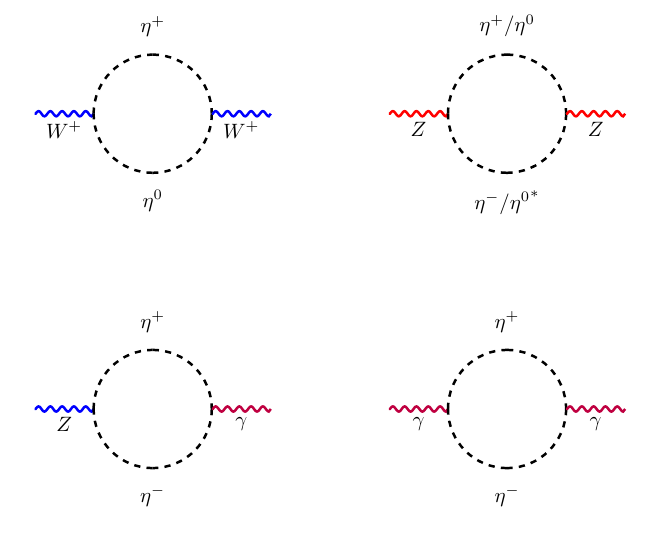}
    \caption{ \begin{footnotesize}One loop polarization diagrams that contribute to the oblique $S$, $T$ and $U$ parameters.                   \end{footnotesize}}
    \label{fig:oneloopmW}
\end{figure}

The presence of a new $SU(2)_L$ doublet dark scalar $\eta$ in our model has important consequences for the $W$ boson mass and the EW precision observables in general. The doublet dark scalar leads to radiative corrections to $W$ boson mass as shown in Fig.~\ref{fig:oneloopmW} top left diagram and hence $W$ boson mass can be increased from SM prediction to that obtained by the CDF-II collaboration. However, as discussed in the introduction, an extension of the electroweak sector of the SM will have an impact on many observables which are constrained by multiple experiments. There is a danger that some of the observables can now deviate from the SM prediction by an unacceptably large amount. As shown by \cite{Peskin:1991sw}, it is convenient to parameterize these new physics effects in terms of the oblique  $S$, $T$ and $U$ parameters. A global fit value of the $S$, $T$ and $U$ parameters can then be obtained by using the EW precision observables as input parameters \cite{ParticleDataGroup:2020ssz}. 
Since the CDF-II results, several groups have now updated the $S$, $T$ and $U$ parameters EW fit \cite{Flacher:2008zq,Asadi:2022xiy,deBlas:2022hdk,Lu:2022bgw,Balkin:2022glu,Paul:2022dds,Gu:2022htv,Kawamura:2022uft}. In this section we show that our model while simple enough has enough richness to change $W$ boson mass to the CDF-II measured value while keeping the oblique parameters within their global 3$\sigma$ range.
%

In the context of our model, only the new scalar doublet $\eta$ contributes directly to the electroweak precision observables via the one-loop polarization diagrams shown in Fig.~\ref{fig:oneloopmW}. The neutral scalar singlet $\xi$ will only have a contribution via the mixing with $\eta^0$, which we take to be small as argued in Sec.~\ref{sec:numass}. Following \cite{Zhang:2006vt} we can calculate the BSM contributions to $S$, $T$ and $U$ of the model as
\begin{eqnarray}
S  & = &  \frac{1}{12 \pi }\log \frac{m_{\eta^0}^2}{m_{\eta^+}^2} \nonumber\\
T & = &  \frac{G_F}{4 \sqrt{2} \pi ^2 \alpha_{em}} \left(\frac{m_{\eta^0}^2+m_{\eta^+}^2}{2}-\frac{m_{\eta^0}^2 m_{\eta^+}^2 }{m_{\eta^+}^2-m_{\eta^0}^2}\log \frac{m_{\eta^+}^2}{m_{\eta^0}^2}\right) \\
U &= &\frac{1}{12 \pi } \left(\frac{(m_{\eta^0}^2+m_{\eta^+}^2) \left(m_{\eta^0}^4-4 m_{\eta^0}^2 m_{\eta^+}^2+m_{\eta^+}^4\right) \log
   \left(\frac{m_{\eta^+}^2}{m_{\eta^0}^2}\right)}{(m_{\eta^+}^2-m_{\eta^0}^2)^3}-\frac{5 m_{\eta^0}^4-22 m_{\eta^0}^2 m_{\eta^+}^2+5 m_{\eta^+}^4}{3 (m_{\eta^+}^2-m_{\eta^0}^2)^2}\right) \nonumber
   \label{eq:stu}
\end{eqnarray}

where $G_F$ is the Fermi's constant and $\alpha_{em}$ is the Fine-structure constant.

In terms of the oblique $S$, $T$ and $U$ parameters, the corrections to the $W$ boson mass are given by \cite{Peskin:1991sw}.

\begin{equation}
 m^2_W = {m^2_W}^\text{(SM)}+\frac{\alpha_{em}\cos^2{\theta_w}}{\cos^2{\theta_w}-\sin^2{\theta_w}}m_Z^2\left[-\frac{1}{2}S+\cos^2{\theta_w}T+\frac{(\cos^2{\theta_w}-\sin^2{\theta_w})}{4\sin^2{\theta_w}}U\right]
\end{equation}

 where $\theta_w$ is the weak angle, $\alpha_{em}$ is the fine-structure constant and $m_W^\text{(SM)}$ is the Standard Model prediction for $m_W$.

For our simple model, one can analytically simplify \eqref{eq:stu} by noting that $S$ and $U$ only depend on the ratio $x = \dfrac{m_{\eta^0}^2}{m_{\eta^+}^2}$, while $T$ is an a dimensional function of $x$ which we call $T_\text{adim}$ times a scaling prefactor $\frac{G_F}{4 \sqrt{2} \pi^2 \alpha_{em}} m_{\eta^0}^2$.

\begin{equation}
    S \equiv S(x), \qquad T \equiv \frac{G_F}{4 \sqrt{2} \pi^2 \alpha_{em}} m_{\eta^0}^2 \, T_\text{adim} (x), \qquad U \equiv U(x)
\end{equation}

 \begin{figure}[h]
        \includegraphics[height = 6 cm]{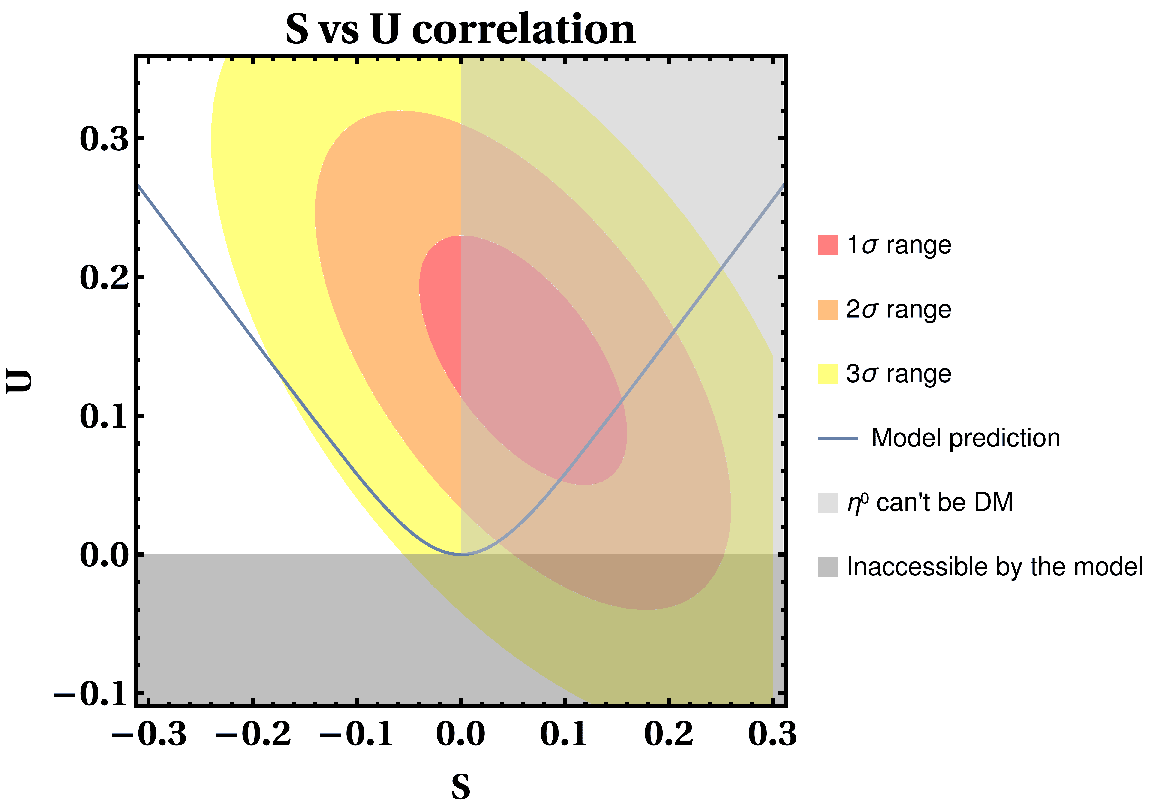} \vspace{0.5cm}\\ 
        \includegraphics[height = 6 cm]{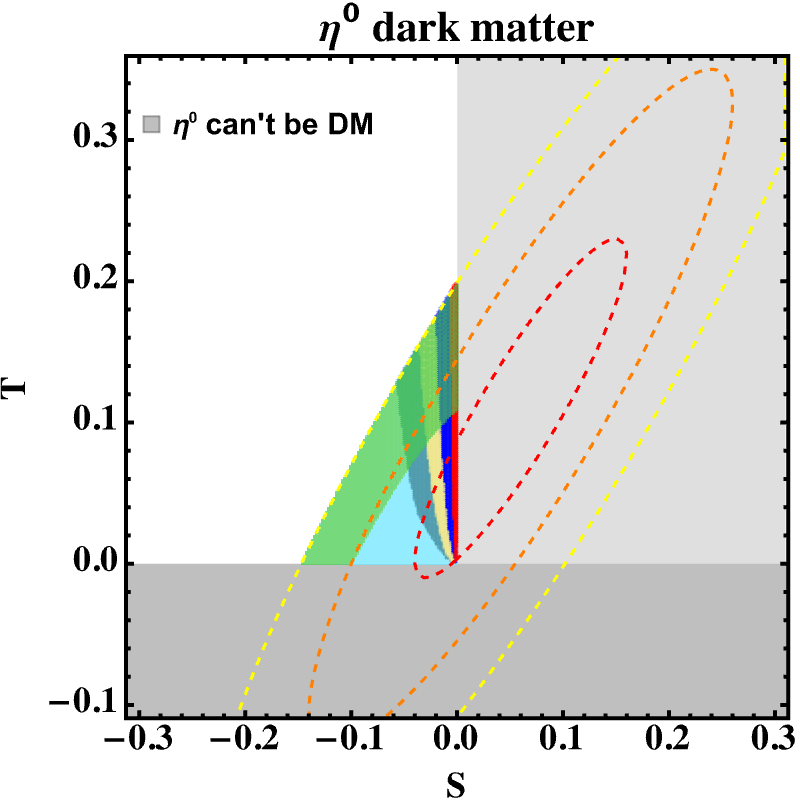}
        \includegraphics[height = 6 cm]{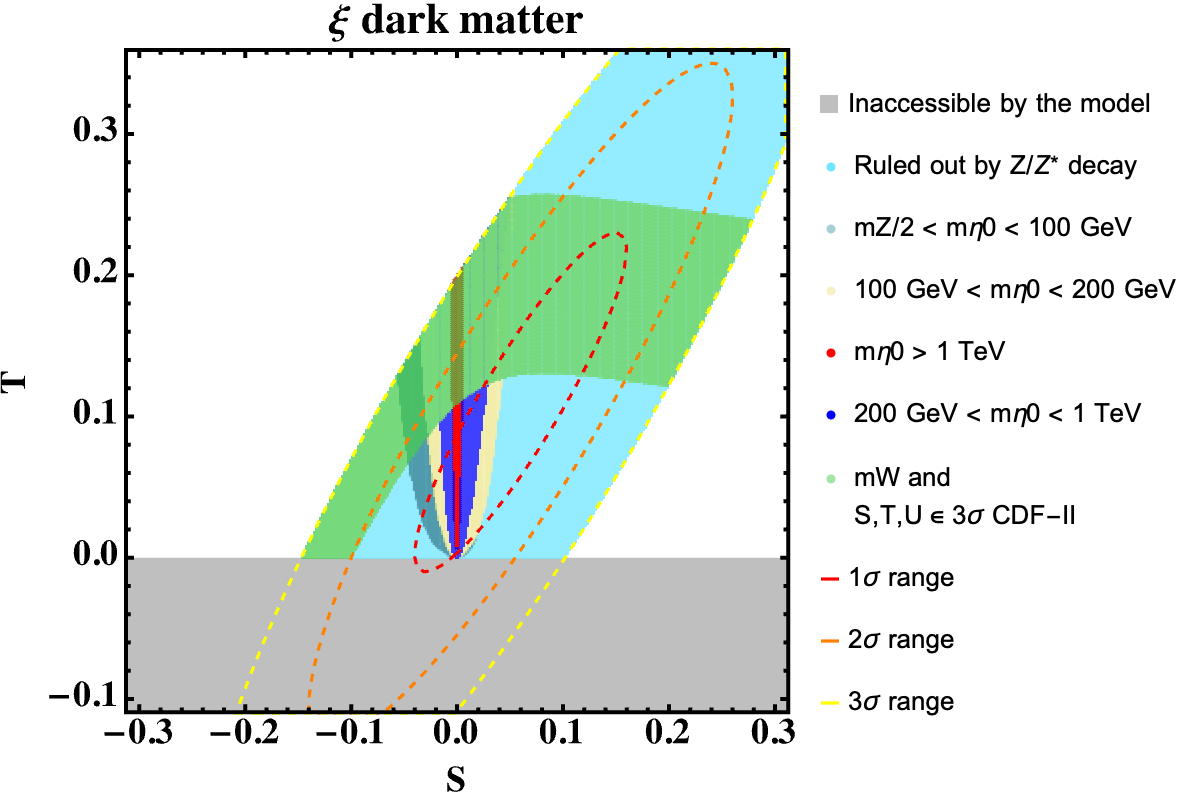}
        \caption{\textbf{Up}: The line represents the correlation between $S$ and $U$ as predicted by the model. \textbf{Down Left:} $S$ vs $T$ correlation with allowed parameter space and constraints for the case of $SU(2)_L$ doublet scalar $\eta^0$ as DM. In this case, the parameter space is tightly constrained. \textbf{Down Right:} $S$ vs $T$ correlation  when the $SU(2)_L$ singlet scalar is DM.
        In both cases, the green band fits the CDF-II $m_W$ measurement at $3\sigma$, while the color code represents the mass of $\eta^0$. It is important to mention that our model cannot accommodate $U<0$ or $T<0$ while $S>0$ implies $m_{\eta^0} > m_{\eta^+}$ (see grey regions).  Furthermore, the $m_{\eta^0} < m_Z/2$ and $m_{\eta^+} \lesssim 70$ GeV regions are ruled out by LEP \cite{Belyaev:2016lok}, shown in cyan color.  %
        Here we are using the EW fits of \cite{Lu:2022bgw}.} 
        \label{fig:SvUvT}
\end{figure}

We can extract some immediate conclusions. Since $S$ and $U$ only depend on the mass ratio $x$, they are perfectly correlated with each other, see the first panel of  Fig.~\ref{fig:SvUvT}. 
The first thing we see is that in our model the oblique parameters $T$ and $U$ cannot be negative as shown in Fig. \ref{fig:SvUvT}. Furthermore, if $\eta^0$ is the DM candidate this means that $m_{\eta^0}^2 < m_{\eta^+}^2$ implying that $x<1$ and therefore $S\leq 0$, $T \geq 0$ and $U \geq 0$. Given the electroweak precision fits \cite{Flacher:2008zq,Asadi:2022xiy,deBlas:2022hdk,Lu:2022bgw,Balkin:2022glu,Paul:2022dds,Gu:2022htv,Kawamura:2022uft}, this is already a severe constraint as shown in the lower left panel of Fig.~\ref{fig:SvUvT}. The requirement  $m_{\eta^0}^2 < m_{\eta^+}^2$ can be accommodated in a $2-3 \sigma$ region of the updated global $S, T, U$ fits, depending on what fit is used for the comparison. On the other hand, if the requirement of $\eta^0$ as DM is dropped then $S$ can be either positive or negative, but still $T, U \geq 0$. This substantially opens up the parameter space as seen in Fig. \ref{fig:SvUvT}. 
Various other constraints like the LEP constraints on Z invisible decay $ Z \to \eta^0 \eta^{0 *}$ and direct search limits from $e^+ e^- \to Z^* \to \eta^+ \eta^-$ further rule out more parts of the parameter space, see Fig. \ref{fig:SvUvT}. Nonetheless for both cases of $SU(2)_L$ doublet scalar $\eta^0$ DM (left panel) as well as for the  $SU(2)_L$ singlet scalar $\xi$ DM (right panel) parts of parameter space survive as shown by the green bands in Fig. \ref{fig:SvUvT}. Thus, regarding precision EW constraints, both $\eta^0$ and $\xi$ as DM are allowed.

\begin{figure}[!h]
        \includegraphics[height = 6cm]{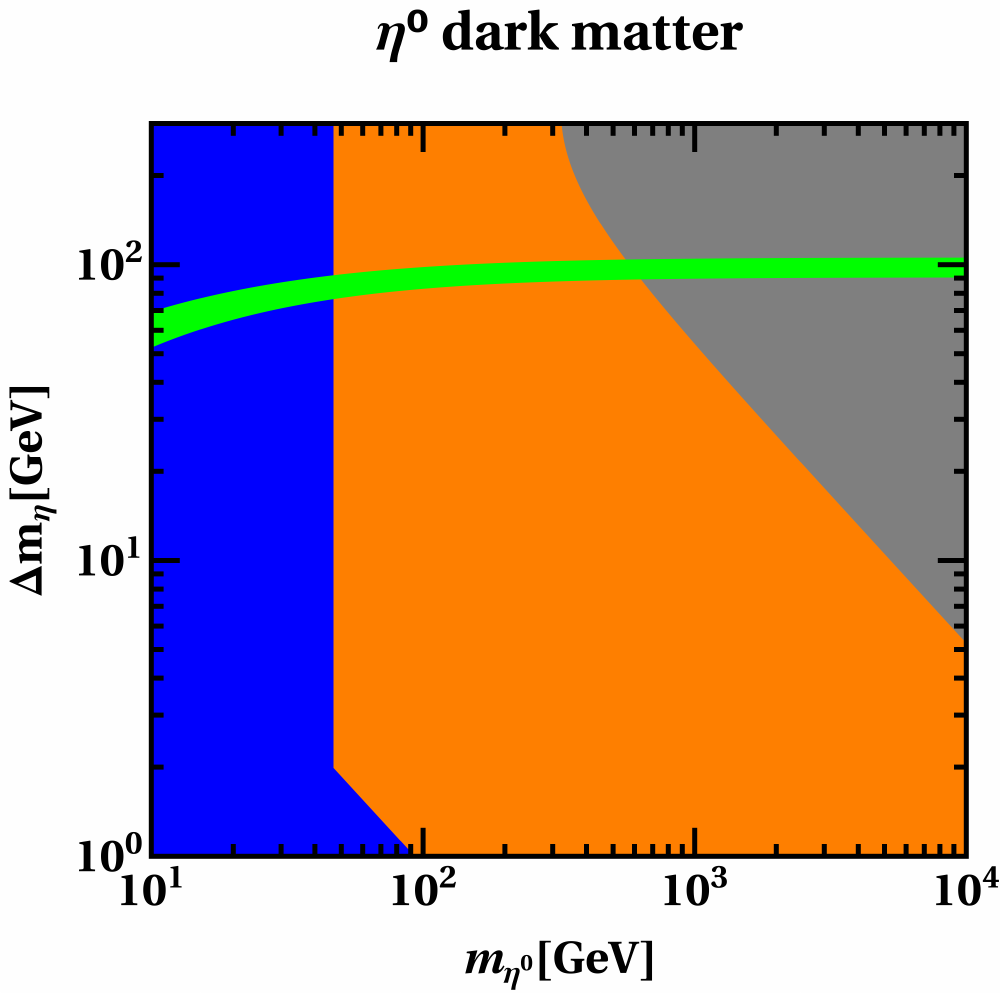}
        \includegraphics[height = 6cm]{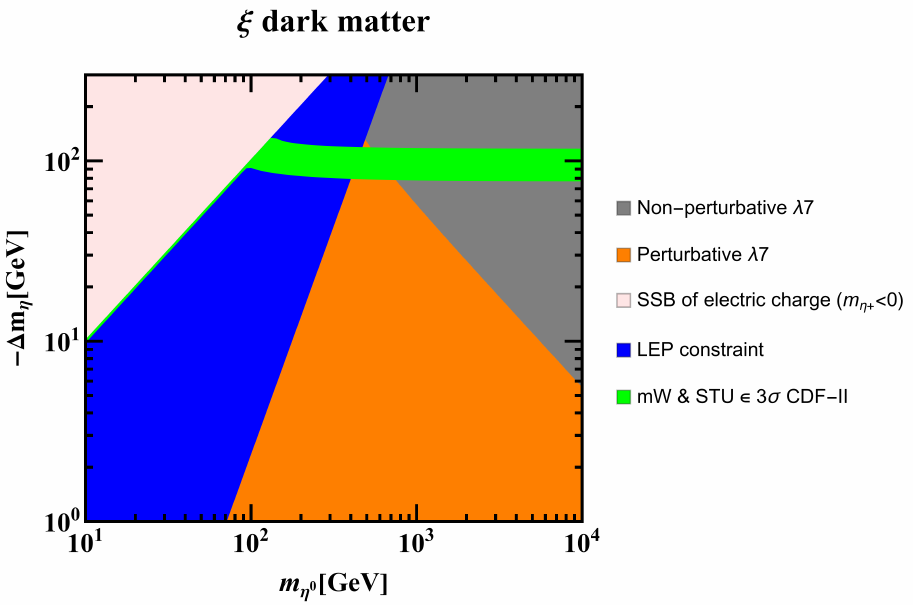}
        \caption{Parameter space in the $m_{\eta^0}$ vs $\pm \Delta m_{\eta}=\pm(m_{\eta^+}-m_{\eta^0})$ plane, in both limiting scalar DM cases. \textbf{Left:} doublet DM, which implies $m_{\eta^0} < m_{\eta^+}$. \textbf{Right:} Singlet DM case and $\Delta m_{\eta}<0$.
        LEP data on $Z$ decays lead to the constraints $m_{\eta^0} > m_Z/2$ and $m_{\eta^+} \gtrsim 70$ GeV \cite{Belyaev:2016lok} through the processes $Z\rightarrow \eta^0 \eta^{0*}$ and $e^+ e^- \rightarrow Z^* \rightarrow \eta^+ \eta^-$. This constraint is shown in blue in both panels. Additionally, the orange region features a perturbative (i.e. $|\lambda_7| < \sqrt 4\pi$) while the grey region features a non-perturbative $\lambda_7$. It is important to mention that $\lambda_7$ is the main parameter responsible for the scalar mass degeneracy $\Delta m_\eta$. Finally, the green regions satisfy the CDF-II $W$ mass measurement in the $3\sigma$ range as well as the $S$, $T$, $U$ $3\sigma$ 2-dimensional regions shown in Fig.~\ref{fig:SvUvT}. }
        \label{fig:etamass}
\end{figure}

The correlation between $m_{\eta^0}$ and $\Delta m_\eta = m_{\eta^+}-m_{\eta^0}$ is shown in Fig.~\ref{fig:etamass} for both cases of $\eta^0$ (left) and $\xi$ (right) being DM. The green region in both panels corresponds to the parameter space where $W$ boson mass is within 3$\sigma$ of the CDF-II range. As can be seen from Fig.~\ref{fig:etamass} part of the green region gets excluded from constraints like perturbativity, spontaneous electric charge breaking vacuum $m_{\eta^+} < 0$, and LEP constraints from  $Z\rightarrow \eta^0 \eta^{0*}$ and $e^+ e^- \rightarrow Z^* \rightarrow \eta^+ \eta^-$. Still, for both cases, a significant part of the allowed parameter space survives all the constraints.


In our numerical scan of the dark sector in Sec.~\ref{sec:DM} we have imposed the $S$, $T$ and $U$ parameters to be inside their $3 \sigma$ 2 dimensional allowed regions. In addition, we also impose the $W$ boson mass to lie inside the recently reported CDF-II $3 \sigma$ value $80.4053 \text{ GeV} \leq m_W \leq 80.4617  \text{ GeV}$ \cite{CDF:2022hxs}.

\section{Dark Matter}
\label{sec:DM}

We now focus on the analysis of the dark sector. In the early universe, all the particles of the dark sector are in thermal equilibrium with the SM fields due to the production-annihilation diagrams shown in the Appendix \ref{sec:appendix}, Figs.~\ref{fig:anihieta} and \ref{fig:anihixi}. As the universe expands, the temperature drops. For unstable particles, there is a maximum temperature below which the thermal bath doesn't have enough energy to produce them, while their annihilation and decay are still allowed and therefore disappear. However, recall that the lightest particle of the dark sector will be completely  stable due to the residual $Z_6$ symmetry. This means that once such stable particle decouples from the thermal bath its relic density will be `freezed out'. This relic density can then be computed and compared with Planck observations \cite{Planck:2018vyg}. This type of DM could also be detected by nuclear recoil experiments such as XENON1T \cite{XENON:2018voc}, see the diagrams shown in Fig.~\ref{fig:detecteta} and \ref{fig:detectxi}.

We already know that the Dirac scotogenic model and its variants can fit all neutrino mixing parameters as well as the observed relic density, as shown in \cite{Guo:2020qin, Hazarika:2022tlc} for all three limiting cases: doublet scalar DM, singlet scalar DM or fermionic DM, as well as in the non-negligible mixing limit. We will now analyze how this situation changes when in addition to these constraints we also impose the electroweak precision fits and the $m_W$ measurement as shown in Sec.~\ref{sec:Wmass}. We performed a detailed numerical scan for the model parameters with various experimental and theoretical constraints. We have implemented the model in SARAH-4.14.5 and SPheno-4.0.5 \cite{Porod:2011nf,Staub:2015kfa} to calculate all the vertices, mass matrices and tadpole equations. In contrast, the thermal component to the DM relic abundance as well as the DM nucleon scattering cross sections is determined by micrOMEGAS-5.2.13 \cite{Belanger:2014vza}.

\subsection{Mainly doublet scalar dark matter}
\label{sec:doubletDM}

As explained before, if the DM particle is mainly formed by the scalar doublet neutral component $\eta^0$ then the ratio $x = m_{\eta^0} / m_{\eta^+} < 0$. This implies that the model prediction for the $S$, $T$ and $U$ oblique parameters satisfy $S \leq 0$, $T, U \geq 0$. Moreover, the allowed parameter space for the masses is also restricted. In addition to the restrictions mentioned above, we have also imposed the following additional conditions when generating the allowed points:

\begin{itemize}
\item Neutrino oscillation parameters as in Sec.~\ref{sec:numass}.
\item Electroweak precision observables and $m_W$ as in Sec.~\ref{sec:Wmass}.
\item Bounded from below scalar potential, ensured by the vacuum stability constraints of Eq.~\ref{eq:vacuumstability}.
\item  Perturbativity of Yukawas and quartic couplings as in Eq.~\ref{eq:perturbativity}.
\item If $\eta^0$ is the DM particle its mass must be smaller than the charged counterpart $\eta^+$. As can be seen from Eqs.~\ref{eq:metaplus} and \ref{eq:metamix}, this implies $\lambda_{7}<0$ in the small mixing limit.
\item The parameters are taken in the ranges shown in Tab.~\ref{tab:parameterrange}.
\item Finally, we impose the LEP constraint on the light-neutral component of a doublet. As shown by \cite{Belyaev:2016lok,Cao:2007rm}, this limit is actually simply $m_{\eta R} + m_{\eta I} > m_Z$ which in our case translates to $m_{\eta^0} > m_Z/2 \approx 45.6$ GeV. In the case of the charged scalar component, this limit is $m_{\eta^+} \gsim 70 \text{  GeV}$.
\end{itemize}

\begin{table}
\begin{center}
\begin{tabular}{|    c   |    c    | c | c |}
  \hline 
  Parameter    &   Range   &   Parameter    &   Range  \\
\hline
$\lambda_{2}$     &  	 $[10^{-8},\sqrt{4\pi}]$            &  $\lambda_{3}$   &  $[10^{-8},\sqrt{4\pi}]$          \\
$|\lambda_{4}|$     &      $[10^{-8},\sqrt{4\pi}]$     &  $|\lambda_8|$   &   $[10^{-8},\sqrt{4\pi}]$   \\
$|\lambda_{6}|$     &      $[10^{-8},\sqrt{4\pi}]$   	 &  $|\lambda_{7}|$   &   $[10^{-8},\sqrt{4\pi}]$   \\
$|\kappa|$          &	     $[10^{-8},30]\text{ GeV}$       &  $\mu^2_{\eta}$  &  $[10^{2},10^{8}]\text{ GeV}^2$\\
$\mu^2_{\xi}$     & $[10^{2},10^{8}]\text{ GeV}^2$ 	 &  $M_{N_i}$ & $[10,10^{5}] \text{ GeV}$  	 \\	
    \hline
  \end{tabular}
\end{center}
\caption{Value range for the numerical parameter scan for S, T and U parameters, relic density and DM direct detection.}
 \label{tab:parameterrange} 
\end{table}

 \begin{figure}[!h]
 \centering
 \textbf{\Large $\eta^0$ dominated DM}\\
        \includegraphics[height=6.5cm]{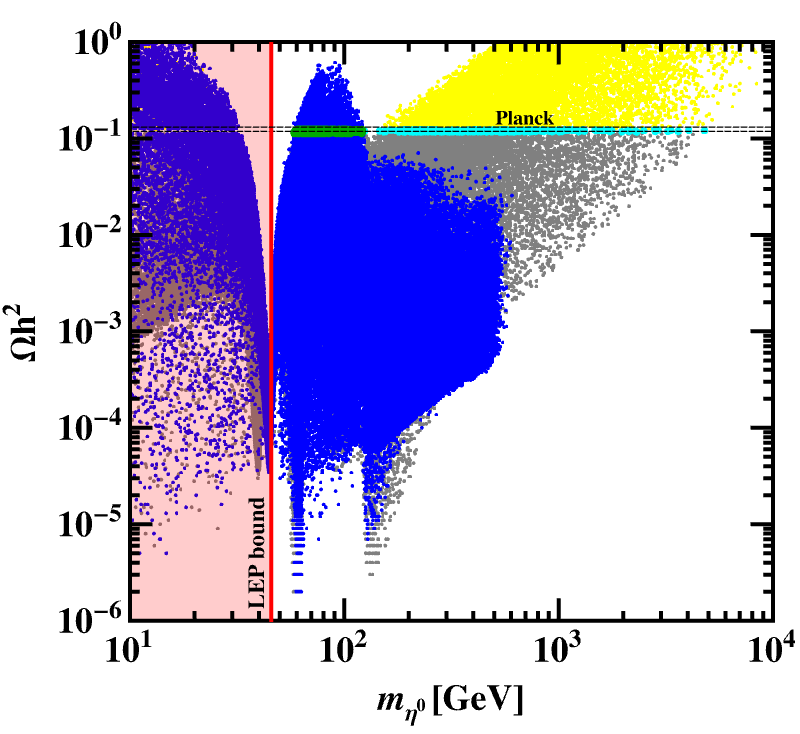}
        \includegraphics[height=6.5cm]{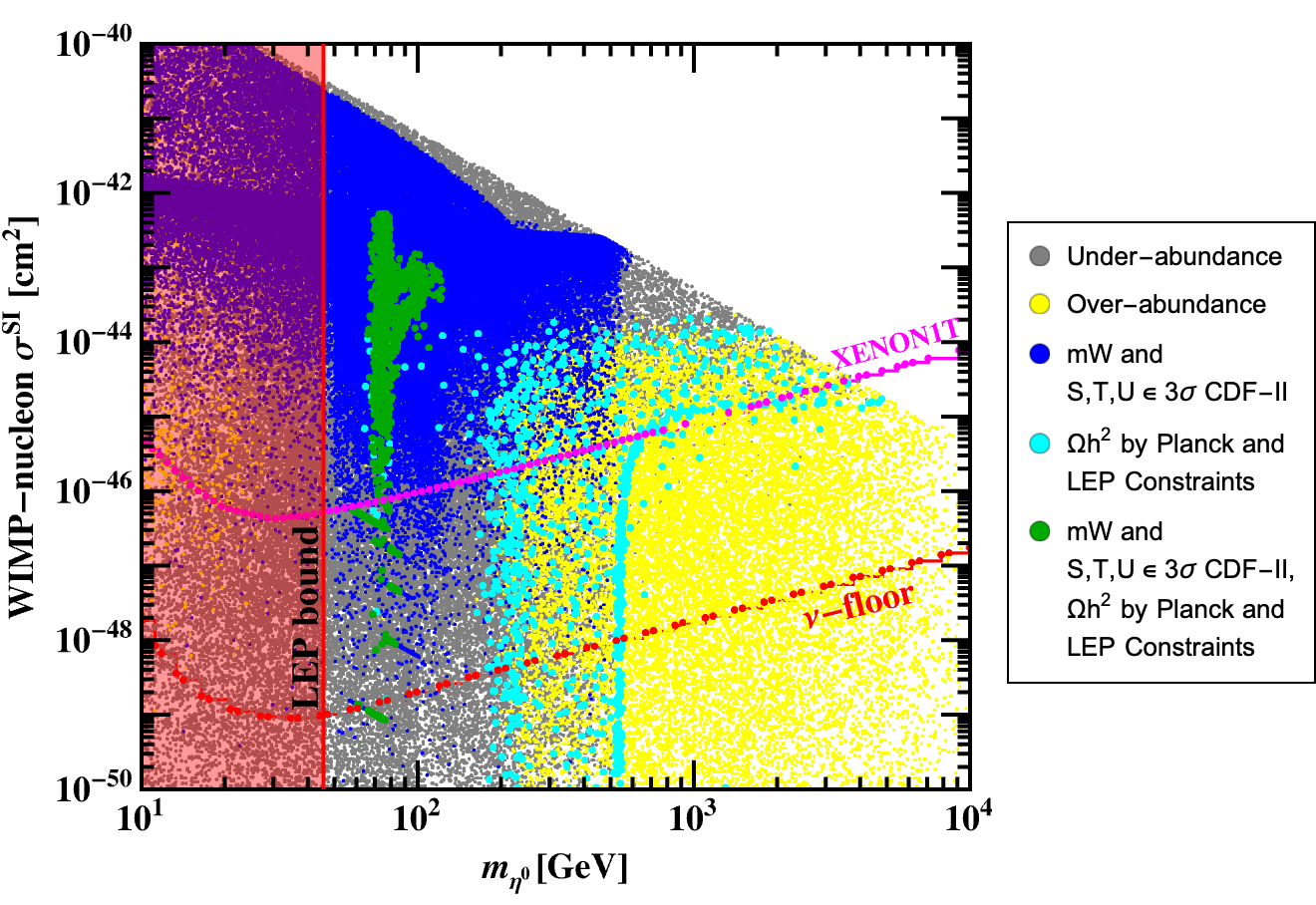}
        \caption{Predictions for the mainly doublet DM case. In both panels, the green points satisfy CDF-II $W$ mass in the $3\sigma$ range, correct relic density and LEP constraints. Yellow/grey points represent over/under abundant relic density \cite{Planck:2018vyg}, respectively, while points with only correct relic density are instead cyan. Blue points satisfy only CDF-II $W$ mass and the $S$, $T$, $U$ parameters in the $3\sigma$ range. The vertical line labelled as `$Z$ decay' is the lower limit to $m_{\eta^0}$ given by LEP. \textbf{Left:} Relic density plot for $\eta^0$ dominated DM. \textbf{Right:} Spin-independent WIMP-nucleon cross section for the $\eta^0$ dominated DM candidate case.}
        \label{fig:doubletDM}
\end{figure}

Fig.~\ref{fig:doubletDM} (left) shows the dependence of the DM relic abundance on the mass of the mainly doublet DM particle, taking into account its annihilation and co-annihilation into bosons and fermions, shown in Fig~\ref{fig:anihieta}. Some of the important features of the relic density plot of DM are: We see the first dip in the relic density plot of DM at around half the mass of the Z boson. This dip is known as the \textquotedblleft Z-boson resonance\textquotedblright or \textquotedblleft Z-pole\textquotedblright. It occurs because, at this mass, the annihilation of doublet DM particles can occur very efficiently through the exchange of a Z boson. As a result, the annihilation cross-section of doublet DM particles is enhanced, leading to a decrease in the relic density of doublet DM at this mass. We found the second dip in the relic density plot of doublet scalar DM when the mass of the scalar DM particle is around half the mass of the Higgs boson. This is because at this mass the annihilation of scalar DM particles can occur very efficiently through the exchange of a Higgs boson. Therefore, the annihilation cross-section of doublet scalar DM particles is enhanced, leading to a decrease in the relic density of DM at this mass. For masses around 90 GeV, quartic interactions between the DM particle and the Standard Model particles become important. This leads to an enhanced annihilation cross-section of the DM particle, particularly into pairs of W bosons. As a result, the effective thermal freeze-out temperature is increased, and the relic density of DM decreases accordingly. This can be observed as a third drop in the relic density plot. Similarly, at a mass of around 125 GeV, the DM particle can annihilate into a pair of Higgs bosons, which becomes an increasingly important annihilation channel due to the enhanced coupling between the Higgs boson and the DM particle. This leads to a decrease in the relic density of DM at this mass, resulting in a fourth drop in the relic density plot.\\

 In Fig.~\ref{fig:doubletDM} (left), all points satisfy the constraints imposed by the Large Hadron Collider (LHC), such as those related to the Higgs boson's invisible decay. In addition, the cyan points in the plot represent the $3 \sigma$ range of the relic density for cold DM, as measured by the Planck collaboration. Our analysis shows that these cyan points cover three mass regions: the low mass region from 10 GeV to around 30 GeV, the medium mass region from 58 GeV to around 122 GeV, and the high mass region from 200 GeV to around 4.8 TeV. The missing mass regions for the correct relic density points correspond to the decrease in the relic density due to the resonant or very efficient annihilation of DM particles into Standard Model (SM) particles through various channels. The grey dots in the plot represent the points where $\eta^0$ is under-abundant. The under-abundance does not mean that $\eta^0$ is ruled out as a DM candidate, it merely means that it cannot be the sole DM candidate and the model needs to be extended to incorporate multi-component DM. Since such multi-component DM requires its own study, we do not go into details about this possibility. The intensity of the grey dots decreases according to the fraction of the relic density they can explain. We observe that for a given DM mass, these grey points correspond to a comparatively larger coupling, leading to small relic density and large WIMP-nucleon cross-sections (Fig.~\ref{fig:doubletDM} right panel) for these grey points. The yellow points correspond to an over-abundance of DM, with a relic density value higher than that observed by the Planck collaboration. As the mass of the DM increases beyond a certain value, the relic density value also increases, explaining the proliferation of these yellow (over-abundance) points for higher DM masses in both plots of Fig.~\ref{fig:doubletDM}. We observe that for a given DM mass these yellow points correspond to a comparatively small coupling, leading to large relic density and small WIMP-nucleon cross-sections (Fig.~\ref{fig:doubletDM} right panel) for these yellow points.\\

 The blue dots indicate the points that satisfy the CDF-II measurement of the W-boson mass as well as the oblique parameter constraints at $3 \sigma$. Since the correction required to $m_W$ is large, these blue points correspond to a relatively small DM mass and relatively large couplings and that is why they mostly lie in the under-abundant region (left panel) and large WIMP-nucleon cross-sections (right panel) region of the plots. Only a small range of low mass  and largish couplings in the 58 GeV to 122 GeV, shown in green points, escapes all the experimental constraints and simultaneously has the right relic abundance. 
 We observe a sudden disappearance of blue points after a DM mass of around 500 GeV. This is because for heavier masses the coupling has to be larger in order to produce the required large correction to $m_W$. But due to the perturbativity constraints, the coupling cannot be increased indefinitely to compensate for increasing mass. This is why we see the cut-off of blue points near 500 GeV.
 The green dots correspond to the points that satisfy the constraints from the LEP and Higgs-invisible searches, in addition to the total relic abundance and the W-boson mass by CDF-II, all at a $3 \sigma$ level of significance. Our analysis indicates that the green points lie in the range of 58 GeV to 122 GeV. This range is bounded by the dips coming from DM mass being half the Higgs mass and it becomes close to the mass of the Higgs.
 Above 122 GeV the relic density decreases and there is a dip at around 125 GeV due to the annihilation of DM particles into a pair of Higgs bosons. While below 58 GeV the DM annihilation to SM particles through gauge boson mediation becomes too efficient (even in the limit of the DM Higgs quartic coupling going to zero) owing to the fact that the gauge bosons go on-shell.\\

 The right side of Fig.~\ref{fig:doubletDM} depicts the relationship between the spin-independent WIMP-nucleon cross-section and the mass of the DM particle. The Xenon-1T experiment has established an upper bound on the nucleon cross-section, and this bound places constraints on WIMP masses over 6 GeV. In our analysis, for DM masses above 86 GeV, the green points on the graph exhibit a higher cross-section than the upper limit established by Xenon-1T. However, for the range of DM masses between 58 and 86 GeV, there are green points that meet theoretical and experimental constraints for mainly doublet scalar DM. Therefore, the only surviving DM masses for this model fall within this range.\\
It is important to mention that the results for the doublet DM case are equivalent to those in the Majorana scotogenic case \cite{Batra:2022pej}. However, in the scotogenic Dirac case, there is an alternative for having scalar DM, namely the singlet DM which we will discuss in Sec.~\ref{sec:singletDM}.

\subsection{Mainly singlet scalar dark matter}
\label{sec:singletDM}
\begin{figure}[!h]
\centering
\textbf{\Large $\xi$ dominated DM}\\
        \includegraphics[height=6.5cm]{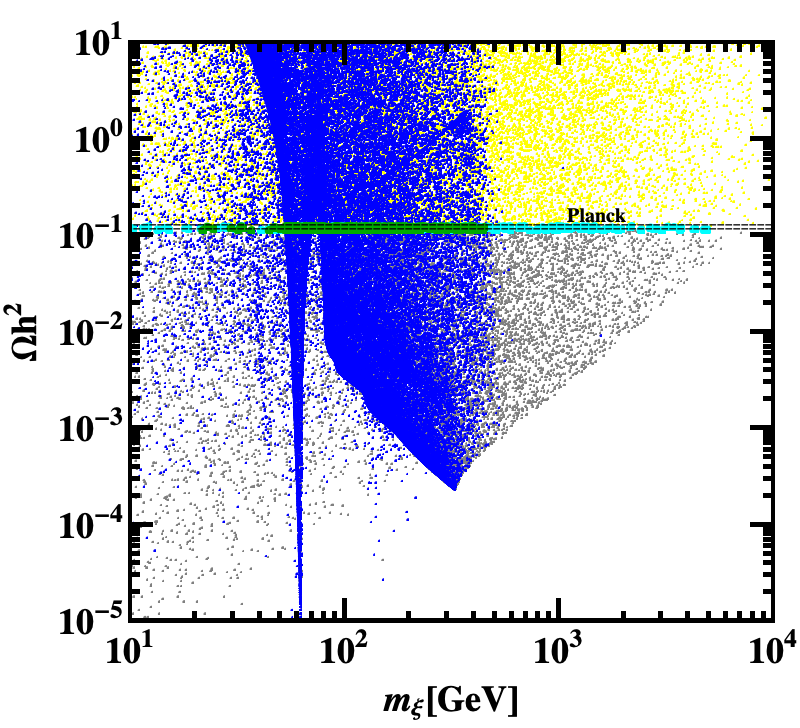}
        \includegraphics[height=6.5cm]{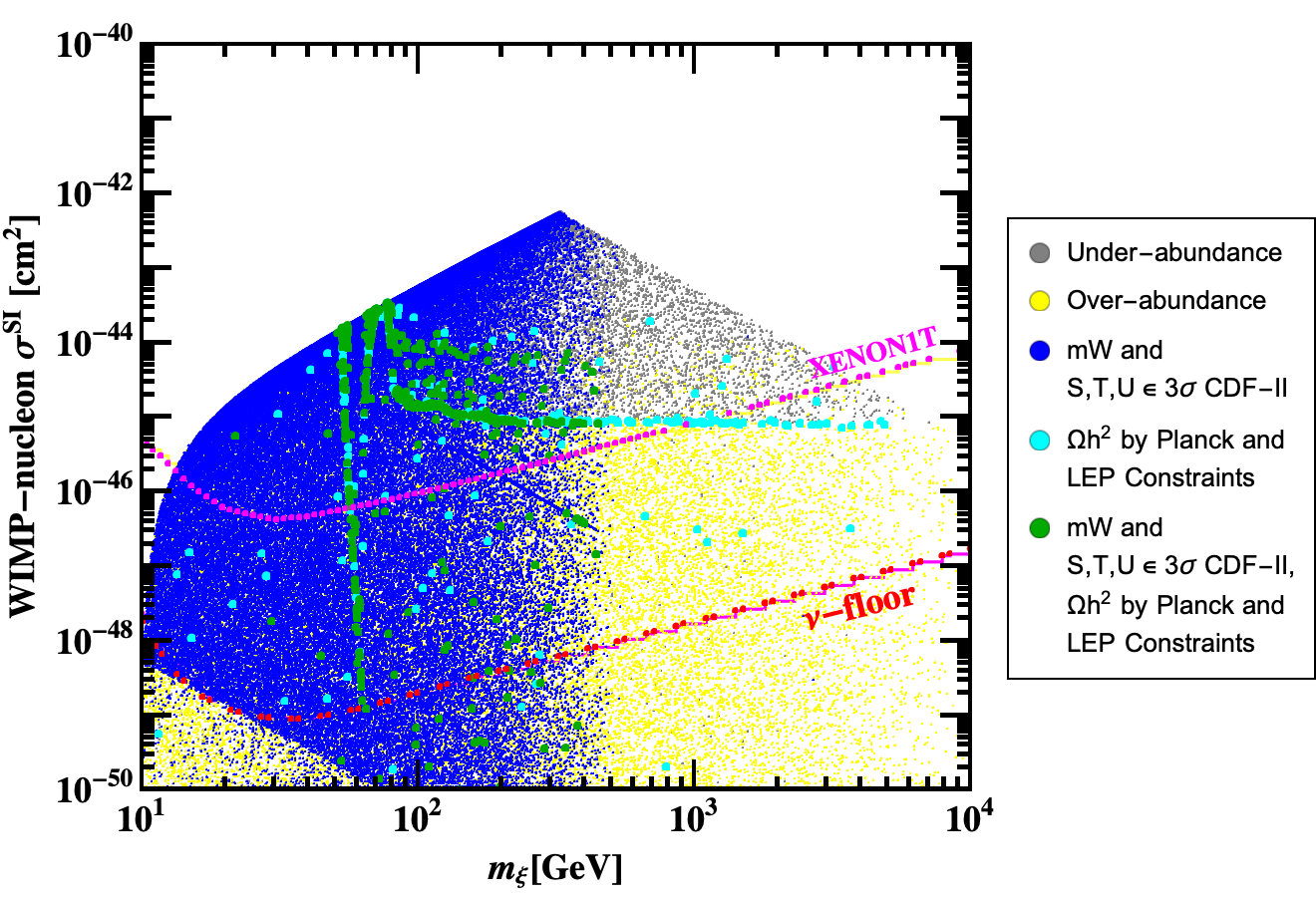}
        \caption{Predictions for the mainly singlet DM case. In both panels, the green points satisfy CDF-II $W$ mass in the $3\sigma$ range, correct relic density and LEP constraints. Yellow/grey points represent over/under abundant relic density \cite{Planck:2018vyg}, respectively, while points with only correct relic density are instead cyan. Blue points satisfy only CDF-II $W$ mass and the $S$, $T$, $U$ parameters in the $3\sigma$ range. \textbf{Left:} Relic density vs singlet DM mass. \textbf{Right:} Spin-independent WIMP-nucleon cross section for the $\xi$ dominated DM candidate vs DM mass. Since the singlet does not participate directly in the $W$ mass it is possible to fit all the experimental constraints.}
        \label{fig:singletDM}
\end{figure}
We now move on to the case in which the DM is mainly formed by the $SU(2)_L$ singlet scalar. If the mainly singlet scalar is lighter than the neutral and charged doublet components it will be the DM particle of the model. It is important to mention that it is now possible to have $S>0$, i.e. $m_{\eta^0} > m_{\eta^+}$, since the charged scalar will now be kinematically allowed to decay into $\xi$ + leptons. This opens the allowed parameter space that can fit $m_W$ and the oblique parameters as shown in Sec.~\ref{sec:Wmass}.

 Moreover, the expressions for $S$, $T$ and $U$ do not depend on the mass of the singlet\footnote{The singlet can participate in the loops only via mixing. Therefore, such contributions will be suppressed by $\sin^2 \theta$, which we are taking small throughout this work. See Sec.~\ref{sec:numass} for a more complete discussion.}. This decoupling between the dark sector and $m_W$ is the main reason why it is now possible to fit the relic density and the EW observables simultaneously.

Fig.~\ref{fig:singletDM} (left) shows the dependence of the DM relic abundance on the mass of the mainly singlet DM particle. Most of the features of this fig. can be understood in an analogous manner as described for the features of Fig.~\ref{fig:doubletDM}. One, however, has to keep in mind that for the current case, the dominant DM annihilation channel is only the Higgs portal as shown in Fig.~\ref{fig:anihixi} while the WIMP-nucleon cross-section is also mostly controlled through the Higgs mediation only as shown in Fig.~\ref{fig:detectxi}.
For the singlet DM case, we observe a clear dip in the relic density plot when the mass of the singlet DM is around half the mass of the Higgs boson. This is because, at this mass, the annihilation of singlet scalar DM particles can occur very efficiently through the exchange of a Higgs boson. As a result, the annihilation cross-section of singlet scalar DM particles is enhanced, leading to a decrease in the relic density of DM at this mass. We observe the direct detection prospects of scalar DM  only through the Higgs portal in the case of a singlet. The mainly singlet DM scenario is relatively less affected by the constraints arising from the CDF-II W mass. However, it is important to mention that if the mainly singlet scalar is considered as the DM particle, then it should have a mass lower than other particles in the dark sector. So, if there is a limit on the mass of the doublet scalar due to constraints from W mass and oblique parameters, then there will also be a limit on the mass of the singlet scalar. that is why the blue points in this case also stop beyond a certain DM mass as for heavier masses, the doublet scalar is too heavy to give desired large corrections to $m_W$ for couplings within the perturbativity limit. Despite this limitation, the singlet scalar can still survive all the constraints for a mass range of up to approximately 500 GeV. This is because the singlet scalar is less constrained by the existing experimental data, compared to the doublet scalar. Therefore, the viable mass range for the singlet scalar is relatively wider, allowing it to evade the constraints and still satisfy the existing experimental bounds.\\

The existing parameter space for both scenarios lies in the relatively low mass and relatively large coupling limits. Thus the prospects of a future collider such as the ILC~\cite{Behnke:2013xla}, CLIC~\cite{CLIC:2018fvx}, FCC-ee~\cite{FCC:2018evy} or CEPC~\cite{CEPCStudyGroup:2018ghi} is very promising. Furthermore, since most of the surviving green points lie above the neutrino floor, the prospects of direct detection in future DM direct detection experiments are also very bright. Finally, improved measurement of $m_W$ can also constrain the surviving parameter space even further. In summary, the model can be tested in different ways in upcoming intensity as well as energy frontier experiments.\\
As a final remark, the fermionic DM case in our model is not  very different from the Majorana scotogenic model studied in \cite{Batra:2022pej}. Moreover, being an SM gauge singlet Dirac fermion, there is no prospect of its  direct detection in currently running or even near future experiments. Thus, the fermionic case at this point is not very interesting from the phenomenological point of view.  Therefore here we have refrained from discussing it in detail.

\section{Conclusions}
\label{sec:conclusions}

We have considered the Dirac scotogenic model presented in \cite{Bonilla:2018ynb} and analyzed its phenomenology in detail. We have shown that the Dirac scotogenic model can reproduce the neutrino masses and mixing, the DM relic abundance and explain the CDF-II $W$ boson mass anomaly in a single framework. Moreover, we find that the case of a mainly scalar doublet DM is constrained for the mass range of 58-86 GeV by the combination of the requirements that $W$ boson mass remains within 3$\sigma$ of the CDF-II measurement and the constraints coming from DM relic density, direct detection and invisible $Z$ boson decays. Unlike the Majorana scotogenic case, here another scalar DM alternative exists namely one where the DM candidate is mainly the $SU(2)_L$ singlet scalar. We showed that if the singlet scalar is the DM candidate then all the above constraints are simultaneously satisfied along with $W$ boson mass within the 3$\sigma$ range of the CDF-II measurement, where the singlet DM mass is constrained upto around 500 GeV.

\section{Acknowledgements}

The authors thank Sudip Jana for the helpful discussions. The work of RS is supported by SERB, Government of India grant SRG/2020/002303. SY acknowledges the funding support by the CSIR NET-JRF fellowship.

\appendix

\section{Annihilation, production and detection of DM}
\label{sec:appendix}

\begin{figure}[!h]
        \includegraphics[width=13cm]{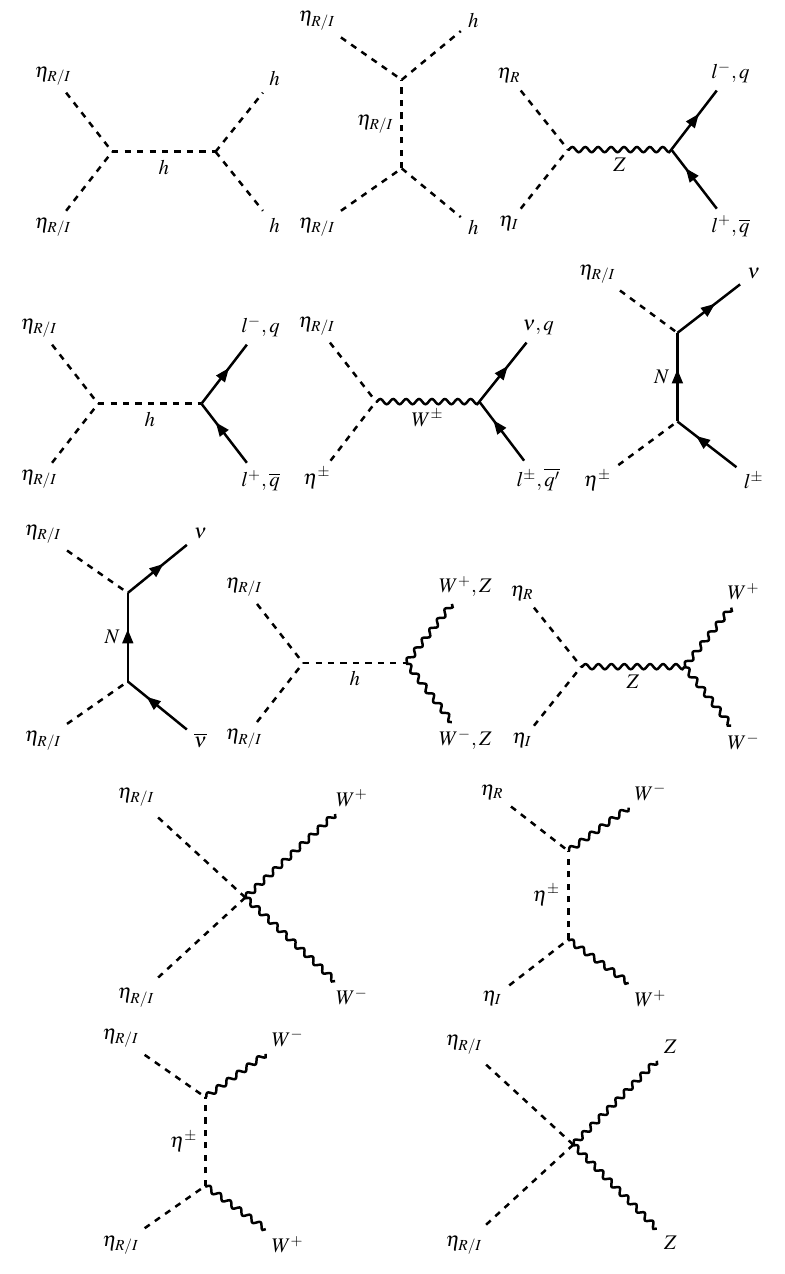}
        \caption{Relevant diagrams for computing the relic density of $\eta^{0}$ dominated DM.}
                \label{fig:anihieta}
\end{figure}

\begin{figure}[!h]
        \includegraphics[height=5cm]{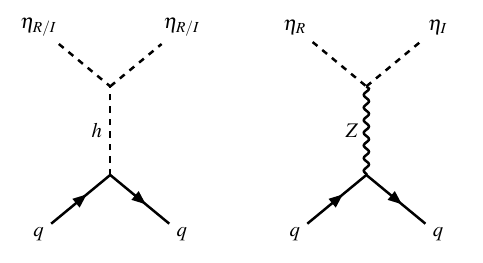}
        \caption{Relevant diagrams for the direct detection of the $\eta^{0}$ dominated DM.}
                \label{fig:detecteta} 
\end{figure}

\begin{figure}[!h]
        \includegraphics[width=13cm]{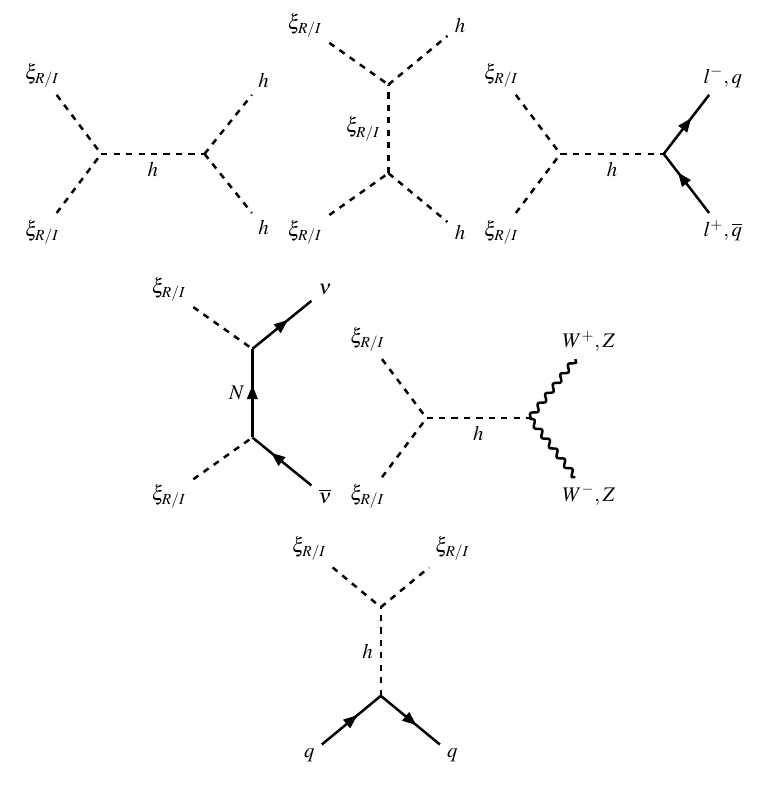}
        \caption{Relevant diagrams for computing the relic density of $\xi$ dominated DM.}
                \label{fig:anihixi}
\end{figure}

\begin{figure}[!h]
        \includegraphics[height=5cm]{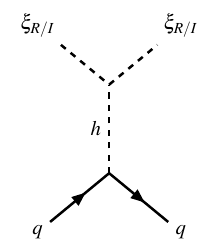}
        \caption{Relevant diagrams for the direct detection of the $\xi$ dominated DM.}
                \label{fig:detectxi} 
\end{figure}

In Figs.~\ref{fig:anihieta} and \ref{fig:anihixi} we list the possible diagrams for production/annihilation of DM, relevant in the early universe, for the cases in which the DM is mainly a doublet or a singlet, respectively. In Figs.~\ref{fig:detecteta} and \ref{fig:detectxi} we show the direct detection prospects of the scalar DM by exchange of a Higgs or $Z$ bosons, in the doublet case, and just a Higgs portal in the singlet case.

\newpage

\bibliographystyle{utphys}
\providecommand{\href}[2]{#2}\begingroup\raggedright\endgroup

\end{document}